\documentclass[11pt]{statsoc} 
\usepackage{amssymb,multicol, amsmath}
\usepackage{xcolor,setspace}

\usepackage{lscape} 
\usepackage[utf8]{inputenc} 

\usepackage[a4paper]{geometry}

\usepackage{amssymb, amsmath}
\usepackage{etoolbox}

\makeatletter
\patchcmd{\@makecaption}
  {\parbox}
  {\advance\@tempdima-\fontdimen2} 
  {}{}
\makeatother  

\usepackage{graphicx}
\usepackage[caption = false]{subfig} 
\usepackage{enumerate}
\usepackage{comment}
\usepackage{natbib, float}

\usepackage{geometry} 
\usepackage{graphicx} 

\usepackage{booktabs}
\usepackage{array} 
\usepackage{paralist}
\usepackage{verbatim}
\usepackage{fancyhdr}
\usepackage{hhline}
\title{Using social contact data to improve the overall effect estimate of a cluster-randomized influenza vaccination program in Senegal}

\author[Gail E. Potter {\it et al.}]{Gail E. Potter}
\address{The Emmes Company, 
Rockville Maryland, 
USA}

\author[Nicole Bohme Carnegie]{Nicole Bohme~Carnegie}
\address{Montana State University, Bozeman Montana, USA}

\author[]{Jonathan D. Sugimoto}
\address{Epidemiologic Research and Information Center, Veterans Affairs Puget Sound Health Care System and Fred Hutchinson Cancer Research Center, Seattle Washington, USA}

\author[]{Aldiouma Diallo}
\address{Institut de Recherche pour le D\'eveloppement, Niakhar Senegal}

\author[]{John C. Victor}
\address{PATH, Seattle Washington, USA}

\author[]{Kathleen Neuzil}
\address{University of Maryland School of Medicine, Baltimore Maryland, USA}

\author[]{M. Elizabeth Halloran}
\address{University of Washington Department of Biostatistics and Fred Hutchinson Cancer Research Center, Seattle Washington, USA}

\begin{document}

\begin{abstract}
This study estimates the overall effect of two  influenza vaccination programs consecutively administered in a cluster-randomized trial in western Senegal over the course of two influenza seasons from 2009-2011.  We apply cutting-edge methodology combining social contact data with infection data to reduce bias in estimation arising from contamination between clusters. Our time-varying estimates reveal a reduction in seasonal influenza from the intervention and a nonsignificant increase in H1N1 pandemic influenza.  We estimate an additive change in overall cumulative incidence (which was 6.13\% in the control arm) of -0.68 percentage points during Year 1 of the study (95\% CI: -2.53, 1.18). When H1N1 pandemic infections were excluded from analysis, the estimated change was -1.45 percentage points and was significant  (95\% CI, -2.81, -0.08).  Because cross-cluster contamination was low (0-3\% of contacts for most villages), an estimator assuming no contamination was only slightly attenuated (-0.65 percentage points).  These findings are encouraging for studies carefully designed to minimize spillover.  Further work is needed to estimate contamination – and its effect on estimation – in a variety of settings.
\end{abstract}


\section{Background}

Influenza is a seasonal respiratory infection that causes a substantial global burden of morbidity and mortality, particularly among children.  One meta-analysis estimated that in 2018 the global burden of influenza among children under 5 was 109.5 million influenza episodes, 870,000 hospital admissions for influenza virus-associated acute lower respiratory infection, and between 13,200 and 97,200 deaths~\citep{wang2020global}. 


In this paper, we use state-of-the-art methodology to estimate the overall effect of annual influenza vaccination of children age 6 months to 10 years ---relative to polio vaccination---on the incidence of influenza in western Senegal. In one analysis of surveillance data in Senegal, 60\% of laboratory-confirmed influenza cases were in children under 5 years of age, and 75\% were in children under 16 years.  While the majority of the population served by the surveillance clinics were children 0-15 years of age, this suggests that a vaccination campaign focusing on young children has the potential to prevent the bulk of influenza case burden on health clinics in Senegal~\citep{niang2012sentinel}.

The study that produced the data analyzed in this paper was a cluster-randomized trial of 20 villages in the Niakhar Demographic Surveillance System (DSS) zone.  Villages were assigned to vaccination of children with either inactivated trivalent influenza vaccine or an inactivated polio vaccine as an active control. There is no national recommendation for routine influenza vaccination in Senegal, hence off-study vaccination was expected to be minimal.  The trivalent influenza vaccine has been shown to be efficacious in reducing influenza infection in children in other settings~\citep{madhi2014influenza, zimmerman20162014}; the Niakhar study was testing the effectiveness of widespread immunization of children to reduce the community burden of influenza.  The primary analysis for this trial analyzed the \emph{total effect} of the intervention~\citep{primary}.  The \emph{total effect} is based on comparing outcomes of treated people in treated villages to those of untreated people in control villages and accounts for protection conferred by receipt of the vaccine as well as from reduction in exposure resulting from vaccination of others in the community.  In this paper, we consider the \textit{overall effect} of the intervention. The \emph{overall effect} is based on comparing the average outcome in treated villages to the average outcome in control villages, so takes into account the effect of the community intervention on both treated and untreated people~\citep{halloran1991direct}.  

The total and overall effects are of interest scientifically because of the presence of interference in infectious disease processes.  Interference ---when one person's treatment can affect another's outcome---is both a boon to disease prevention and a classic inferential problem in infectious disease research.  The benefit: the very nature of the process induces dependence between people's outcomes, and treating one person may prevent another's infection.  The drawback: observations are no longer independent, and most mainstream causal inference tools cannot account for the induced dependence.  The main approach to dealing with interference is to use cluster-randomized trials (CRT), which allow for dependence within cluster.  The assumption of no interference that would be made in a traditional individually-randomized controlled trial is thus weakened to \textit{partial interference}---an assumption of no interference between clusters~\citep{Sobel2006}.  Violation of the partial interference condition is referred to as \emph{contamination}~\citep{HudgensHalloran2008}.

Typical methods for estimating the overall effect assume partial interference (e.g.,~\cite{HalloranStruchiner1991, liu2014large}).  However, for socially contagious outcomes such as infectious diseases, partial interference will not be satisfied if members of treated clusters come in contact with people from untreated clusters (and vice-versa). Recent methodological developments have explored incorporation of measured contamination data into estimation and testing methods to explicitly adjust for interference.   See~\cite{halloran2016dependent} and~\cite{savje2017average} for reviews of recent efforts to develop causal inference methods that account for partial interference as well as more general forms of interference.  Some of these methods incorporate detailed social network structure~\citep{eckles2017design, toulis2013estimation, aronow2017estimating,ugander2013graph}, but such detailed network data is not always available or easy to obtain. In this study, the complete social network was not observed, but information was collected on rates of contacts within and between villages.  Most relevant to this data structure and to our interest in the overall effect is a method developed by~\cite{carnegie2016estimation}.  It is well known that when contamination is present, the overall effect estimate is attenuated.  The authors developed a method to explicitly incorporate measured contamination data into the estimation procedure and demonstrated that this adjustment removes the attenuation of the overall effect estimate.  We apply this method to estimate the overall effect accounting for cross-cluster contamination and compare it to the estimate that would be obtained assuming partial interference.

This paper continues as follows.  In Section \ref{sec:data}, we describe the data; in Section \ref{sec:methods} we describe the causal model and data preparation.  The results of causal effect estimation are given in Section \ref{sec:results}, and implications and limitations are discussed in Section \ref{sec:concl}.

\section{Data Collection}
\label{sec:data}

The data were collected in a cluster-randomized clinical trial conducted in the Niakhar Demographic Surveillance System (DSS) zone from 2009-2011. This study, ClinicalTrials.gov NCT00893906, is closed, and the primary results for the first year of the trial have been published~\citep{primary}.  Among thirty villages in the Niakhar DSS zone, twenty were selected as clusters for inclusion in the trial and randomized in a 1:1 ratio to receive a blinded vaccination campaign of either inactivated trivalent influenza vaccine (TIV) or inactivated poliovirus vaccine (IPV) as an active control.   From here on, villages that received TIV will be referred to as ``treated'' and those that received IPV as ``control''.  The same villages were followed for two influenza seasons (2009-2010 and 2010-2011).  Different formulations of trivalent influenza vaccine were given during the two years; the second formulation included the H1N1 2009 ``swine'' pandemic strain of influenza, but the first formulation did not.  

Within each treatment group the goal was to vaccinate up to 5,000 children 6 months to 10 years of age in the following approximate numbers per age-group: 1,270 children 6-35 months of age; 2,835 children 36 months to 8 years of age; and 895 children 9-10 years of age.  Vaccinees received age-specific doses.  In villages assigned to receive influenza vaccine, 3,906 (78.1\% of target number for vaccination) were vaccinated with Dose 1, while 3,843 (76.9\% of the target) of those in control villages were.  These numbers comprised 66.6\% and 66.2\% of age-eligible children, respectively.

The primary outcome of the study was laboratory-confirmed symptomatic influenza infection.  A combination of active and passive surveillance was used for the primary outcome in the Niakhar DSS zone.  In this geographic area, residences are organized in compounds, clusters of dwellings typically housing an extended family.  For the twenty villages randomized in the study, field workers visited compounds on a weekly basis to inquire about the occurrence of influenza symptoms.  If the person had experienced influenza-like illness (defined as fever or feverishness, cough, sore throat, nasal congestion, and/or rhinorrhea) in the past 7 days, then the field worker consented them into the surveillance study and documented symptoms and epidemiologic data.  Cases of influenza-like illness were reported to the study center, and nasal and throat swab specimens were collected.  In addition, individuals seeking medical care at any of the three Niakhar DSS health posts at any time throughout the year were assessed by health post medical staff or a study physician to determine if the person had influenza-like illness.  These individuals were consented into the surveillance study, their symptoms were documented, and nasal and throat swab specimens were obtained for influenza testing.   

When individuals with influenza-like illness enrolled into the surveillance study, they also responded to a survey about their travel and social contact patterns during the prior three days.  For each day, the respondent provided the number of people she contacted in her own compound in the morning and the afternoon/evening.  In addition, she indicated yes or no to whether she had visited a list of locations: another compound (up to five could be identified in the survey), a market, mosque or church, field, school, sports field or public place, outside the study zone, or another location.  For each location visited, the village identification code (and compound identification number, where applicable), the time of day visited (AM, PM, or both), and the number of persons the respondent spoke with during the visit were recorded.  For additional details, refer to the example survey form in the Appendix.

Village of residence was recorded during  quarterly censuses conducted by the Niakhar DSS~\citep{delaunay2002niakhar, delaunay2013profile}. If participants moved during the trial, their departure date, arrival date, and village of their new residence were recorded.  Those who moved a second time had their departure date (but not residence after second move) recorded as well.  The cleaning that was performed after receiving the residence data from the DSS is described in the Appendix. 

\section{Analytic Methods}
\label{sec:methods}

\subsection{Causal effect estimation}
In this paper, we consider two estimators for the overall effect of influenza vaccination relative to polio vaccination.  The first estimator assumes partial interference (i.e., no contamination), and we refer to it as the \textit{no-contamination estimator}.  The second explicitly accounts for interference generated by contacts to villages of the opposite treatment assignment; we refer to this as the \textit{contamination-adjusted estimator}.  

To account for contamination, we use the method developed in~\cite{carnegie2016estimation}. This approach uses an additive hazards model~\citep{Aalen1989} for the time to first event but includes a modified treatment variable to account for contacts occurring between clusters in a cluster-randomized trial.  Typically, the treatment variable $Z$ is a binary indicator such that $Z=1$ for participants from treated villages and $Z=0$ for those from control villages. This is the treatment variable used to calculate the no-contamination estimator.  To account for interference between clusters, we use an alternate treatment variable $M$, which is the proportion of contacts of residents of the participant's village that are with treated villages.  It can be thought of as a village-level intensity of exposure to the treatment conditions, and will range from 0 (if all contacts reported in a village are with control villages) to 1 (if all of the contacts reported in a village are with treated villages).  Note that if no contamination is present, then this modified treatment variable reduces to the binary treatment variable used to calculate the no-contamination estimator.

The additive hazards model used to obtain the no-contamination estimator for an individual in cluster $j$ is 
\[
\lambda_j(t|Z) = \beta_0(t) + \beta_Z(t)z_j,
\]
where $z_j$ is a binary treatment indicator for  cluster $j$.  The contamination-adjusted estimator is obtained from the following model for individual in cluster $j$:
\[
\lambda_j(t|M) = \beta_0(t) + \beta_M(t)m_j,
\]
where $m_j$ is the total percentage of contacts of susceptibles in cluster $j$ that are with treated clusters.  Note that $m_j$ is a cluster-level variable, but the model is an individual-level model, with individuals in the same cluster taking the same value for $m_j$.

The coefficient of interest in the additive hazards model---corresponding to the treatment variable---is potentially time-varying.  For this reason, we report both that coefficient (visually) and the difference in cumulative hazard of influenza due to the treatment.  Because the cumulative hazard is low, this is approximately equal to the difference in cumulative incidence due to treatment. The time-varying coefficients are visualized by displaying the value of their integrals, $\int_0^t \beta_Z(t)dt$ and $\int_0^t \beta_M(t)dt$, as a function of time.  These integrals represent the cumulative hazard difference over the time interval [0,t] and are estimated using the nonparametric approach proposed by~\cite{Aalen1989}.  Since the nonparametric estimation approach (based on step functions) produces curves that are not always differentiable, the additive treatment effect is not explicitly estimated, but it is visualized as the slope of the curve~\citep{Aalen1989}.  Estimation is implemented with the \texttt{aalen} function in the R package \texttt{timereg} to fit the additive hazards models~\citep{Rtimereg, r2017}, and the {R} code used is provided in the Appendix.  The effects are displayed together with confidence intervals based on robust (sandwich) standard errors which take into account the clustering; these are also provided by the \texttt{aalen} function. 

The estimand of interest, which we will denote $\beta(t)$, is the population-averaged difference in hazard of laboratory-confirmed symptomatic influenza infection associated with a change from 0\% to 100\% exposure to treatment.  While $\hat{\beta}_Z(t)$ is a consistent estimator for $\beta(t)$ in the absence of contamination, \cite{carnegie2016estimation} proved that $\hat{\beta}_M(t)$ is a consistent estimator for $\beta(t)$ in the presence of measured contamination.

This additive hazards model for interference has a natural correspondence to a deterministic compartmental model such as an SIR model (Susceptible-Infectious-Recovered; see, e.g., \cite{KeelingRohani2008}).  This relationship results from the assumption of the compartmental model that the transmission rate is a product of the contact rate and the per-contact transmission probability.  We provide further details on this relationship in the Appendix. This correspondence supports application of our method to influenza, which is frequently modelled with an SIR or SEIR (Susceptible-Exposed-Infectious-Recovered)  model~\citep{coburn2009modeling}.  Since the length of the exposure state is irrelevant to modelling disease-free survival, this method gives identical results under SIR and SEIR assumptions~\citep{carnegie2016estimation}.

While Cox regression is frequently used for survival analysis, the Cox proportional hazards model does not share this natural correspondence to epidemic compartmental models.  Another advantage that the additive hazards model has over the proportional hazards model is collapsibility, which implies that the treatment effect is the causal effect of interest whether or not covariates are included in the model.


Analyses were performed separately for Year 1 and Year 2 of the study.  Inputs to the additive hazards model are the time to event (or censoring) for each person, infection status, and the percentage of contacts to treated clusters.  Calculation of time-to-event for each survey year is described in detail in the Appendix. One irregularity in data collection is noteworthy: during Year 2 of the study, household surveillance was not performed during a strike of field workers that lasted from Jan 3, 2011 through Feb 18, 2011.  This could introduce bias since the rate of reporting infections during household visits (as opposed to health posts) was higher in treated than control villages (87.5\% and 83\%, respectively).  To prevent such bias, we analyzed a shorter time interval for Year 2 by censoring observations at the start of the strike.  The full Year 2 estimates are included as a secondary analysis.

\subsection{Calculation of treatment exposure estimates}
\label{subsection:interference}

The treatment exposure value for village $j$ is the proportion of contacts that susceptible people in village $j$ made with people in treated clusters.  For control villages, this variable is the percentage of contacts to treated villages (the contamination estimate itself).  For treated villages, however, the treatment exposure value is one minus the percentage of contacts to people in control villages (i.e., one minus the contamination estimate).  

The contact survey defined a ``contact'' as a conversation occurring between two people in the same location.  The contact survey collected numbers of contacts in various locations at two time points (AM and PM) for three consecutive days: the survey day and the two prior days.  Numbers of contacts recorded on the survey day are subject to truncation bias because most surveys were administered in the morning and exclude contacts occurring after the time of the survey. Contact patterns for asymptomatic participants are included in the data since some participant's symptoms began on the day of or the day before the survey.  We analyze only data collected from the time point two days before the survey date because this time point includes more reports from asymptomatic people.  Additionally, a social network analysis of these data found no difference in numbers of contacts recorded the day before the survey vs. two days prior -- so there is no evidence that the earlier time point is subject to recall bias~\citep{potter2019networks}. 

The survey did not elicit how many of the morning contacts were repeated in the afternoon/evening.  We analyze contacts reported in the morning as treatment exposure rates were similar between morning and afternoon contacts (Appendix Table~\ref{tab:am_pm}).

Our treatment exposure estimates take into account the percentage of contacts reported while the respondent was visiting treated villages (Section~\ref{subsection:village}) and the percentage of contacts reported in the respondent's own home (compound) that occurred to visitors from  treated villages (Section~\ref{subsection:visitors}).  

\subsubsection{Percentage of contacts in treated villages}
\label{subsection:village}

For each village, we calculated the percentage of contacts reported while respondents from that village were located in treated villages.  The denominator was the sum of contacts reported by village residents; the numerator was the sum of those contacts whose reported location was a treated village.  Contacts reported to villages that are not in the trial were included in the denominator and are treated the same as contacts to control villages. The numerator included contacts reported in the respondent's own compound if the respondent was a resident of a treated village.  For participants who moved mid-study, the village of residence is the reported village of residence at the time of the contact survey.

We initially calculated treatment exposure rates using reports by asymptomatic people only, assuming that this would be more representative of behavior when uninfected and that the symptomatic people would travel less. We compared these to the estimates based on reports by symptomatic people and (counterintuitively) found that symptomatic reports included slightly higher rates of contacts to clusters of the opposite treatment assignment (Appendix Tables~\ref{tab:symp_asymp} and~\ref{tab:symp_asymp_imp}).  This is likely because cross-cluster contact rates are fairly low overall and because less data is available for asymptomatic reports, so the small amount of data from  asymptomatic respondents includes fewer non-zero counts.  Therefore we combined data from both asymptomatic and symptomatic people to estimate the treatment exposure variable more precisely.

\subsubsection{Incorporating treatment exposure from visitors to the respondent's compound}
\label{subsection:visitors}
The above approach assumes that the location of a contact reported by the respondent indicates the residence of the person contacted. As such it does not account for visitors to one's compound from a cluster of the opposite treatment assignment, so may underestimate cross-cluster exposure. To incorporate exposure from visitors into the estimate, we will define some notation and first consider the estimates for people living in control clusters. Suppose there are $n_j$ people living in cluster $j$, and let $D_i$ denote the number of contacts reported by person $i$ who lives in cluster $j$. Let $T_i$ denote the number of contacts person $i$ has made in a location in a treated cluster. Let $p_j$ denote the proportion of contacts in cluster $j$ to people from treated clusters. We have estimated this as
$$\hat{p_j}=\frac{\sum_{i=1}^{n_j}T_i }{\sum_{i=1}^{n_j}D_i}$$
We need to update the numerator to include contacts occurring within the respondent's own compound to visitors from other clusters. We can use estimates reported by these visitors, rather than by respondents in cluster $j$, to obtain this information. Let $V_{trt,j}$ denote the total number of contacts reported by people in any treated cluster during their visits to compounds in cluster $j$.  While these contacts contribute to the denominator in the above estimator, they do not contribute to the numerator (because they occurred within the respondent's assigned cluster), but should.  Therefore, when $j$ is a control cluster, our updated estimate incorporating this exposure is:
$$\hat{p_j}=\frac{\sum_{i=1}^{n_j}T_i +V_{trt,j}}{\sum_{i=1}^{n_j}D_i} = \frac{\sum_{i=1}^{n_j}T_i}{\sum_{i=1}^{n_j}D_i}+\frac{V_{trt,j}}{\sum_{i=1}^{n_j}D_i}$$
The rationale for this adjustment is explained in  detail in~\cite{potter2019networks}, and an explanation tailored to this setting is provided in the Appendix.  

An analogous update is needed for residents of treated clusters.   For these respondents we need to account for visits from members of control clusters. Letting $V_{ctr,j}$ denote the total number of contacts reported by people in any control cluster during their visits to compounds in cluster $j$. When $j$ is a treated cluster, our updated estimate incorporating this exposure is:
$$\hat{p_j}=\frac{\sum_{i=1}^{n_j}T_i -V_{ctr,j}}{\sum_{i=1}^{n_j}D_i}=\frac{\sum_{i=1}^{n_j}T_i}{\sum_{i=1}^{n_j}D_i}-\frac{V_{ctr,j}}{\sum_{i=1}^{n_j}D_i}$$

\subsection{Multiple Imputation for Missing Contact Data}
\label{subsection:imputation}

The submitted contact surveys had a large number of missing fields, which, if not modelled appropriately, could create bias in the estimates of cross-cluster exposure.  For locations visited outside the home two days before the survey, 24\% are missing time of day, 59\% are missing the number of people contacted, and 32\% do not have a village number recorded.  The survey design elicited at-home contacts differently than those that occurred outside the home: the numbers contacted at home in the morning and in the afternoon/evening were recorded, so village and time point were not collected as separate variables.  Furthermore, in 60\% of analyzed surveys, the number contacted at home in the morning was missing.  

We used multiple imputation, expanding on the procedure used in another analysis of this data set~\citep{potter2019networks} to adjust for missing contact data. For outside-home locations, up to four variables may be missing: the response to ``Was this location visited?'', the time of day (AM or PM) the location was visited, the number of people contacted at that location, and the village where the location is located. The responses to whether the location was visited were imputed based on a log binomial regression model with location type, symptom status, and age category as predictors, stratified on day relative to the survey day. Missing times were imputed by sampling from the distribution of non-missing times for that location type. To impute missing numbers of contacts for non-home locations, we fit a negative binomial distribution to the reported contact numbers, predicting the number contacted by the location, symptom status, time of day, and age category. For at-home contacts, we predicted number contacted based on symptom status, time of day, day relative to survey day, age category, and gender. Missing villages for out-of-home contacts were sampled from the observed distribution of visited villages for the respondent's village of residence, combining data from both survey days.  As such we are assuming the data are missing at random; in other words, the predictors in our imputation model are sufficient to explain the distribution of unobserved values~\citep{rubin1976inference}.

We created twenty imputed data sets, calculated percentages of contacts to treated clusters for each village in each of these imputed data sets, and combined the percentages using standard rules for combining multiply imputed data~\citep{rubin1987}. 

\section{Results}
\label{sec:results}

Table~\ref{tab:interference} displays the treatment exposure estimates for each village enrolled in the trial based on the multiply imputed data.  For each village, we display the percentage of contacts reported when the respondent visited treated villages, the estimated percentages of contacts from visitors from villages of the opposite treatment assignment, as well as the overall percent of contacts to treated villages, which was used as a covariate in the contamination-adjusted model.  The overall percentages are generally close to zero for control villages and close to 100 for treated villages, with a few exceptions.

\begin{table}
\caption{\label{tab:interference}Percentages of contacts with residents of treated clusters based on (1) contacts reported while located in treated clusters, (2) contacts in the respondent's own compound to visitors from clusters of the opposite treatment assignment, and (3) total percentages of contacts to residents of treated clusters (treatment exposure).}
\centering
\begin{tabular}{lcccc}
 \hline
 & Treatment & Percent reported & Percent  &  Treatment \\ 
Village & Assignment &  in treated clusters &  from visitors & exposure \\ 
  \hline
Kalome Ndofane & Vaccine & 100 & 0 & 100 \\ 
  Ngayokheme & Vaccine & 99 & 0 & 99 \\ 
  Ndokh & Vaccine & 99 & 1 & 99 \\ 
  Ngangarlame & Vaccine & 99 & 0 & 99 \\ 
  Diohine & Vaccine & 99 & 0 & 98 \\ 
  Mokane Ngouye & Vaccine & 99 & 1 & 98 \\ 
  Nghonine & Vaccine & 98 & 2 & 96 \\ 
  Logdir & Vaccine & 95 & 2 & 93 \\ 
  Darou & Vaccine & 96 & 5 & 90 \\ 
  Poudaye & Vaccine & 93 & 2 & 90 \\ 
  \hline
  Ngalagne Kop & Control & 0 & 0 & 0 \\ 
  Mboyene & Control & 0 & 0 & 0 \\ 
  Poultok Diohine & Control & 0 & 0 & 0 \\ 
  Bary Ndondol & Control & 0 & 1 & 1 \\ 
  Toucar & Control & 1 & 0 & 1 \\ 
  Gadiak & Control & 2 & 0 & 2 \\ 
  Godel & Control & 2 & 0 & 2 \\ 
  Khassous & Control & 3 & 0 & 3 \\ 
  Kothioh & Control & 3 & 0 & 3 \\ 
  Meme & Control & 14 & 0 & 14 \\   
  \hline
\end{tabular}
\end{table}

\begin{figure}
\caption{\label{fig:year1}
Estimated effects of the influenza vaccination program for Year 1 (July 2009 - May 2010) of the study. The time-varying effects are the difference in cumulative incidence of lab-confirmed symptomatic influenza infection between groups, measured in percentage points.}
\centering
\includegraphics[width = 1.1\textwidth]{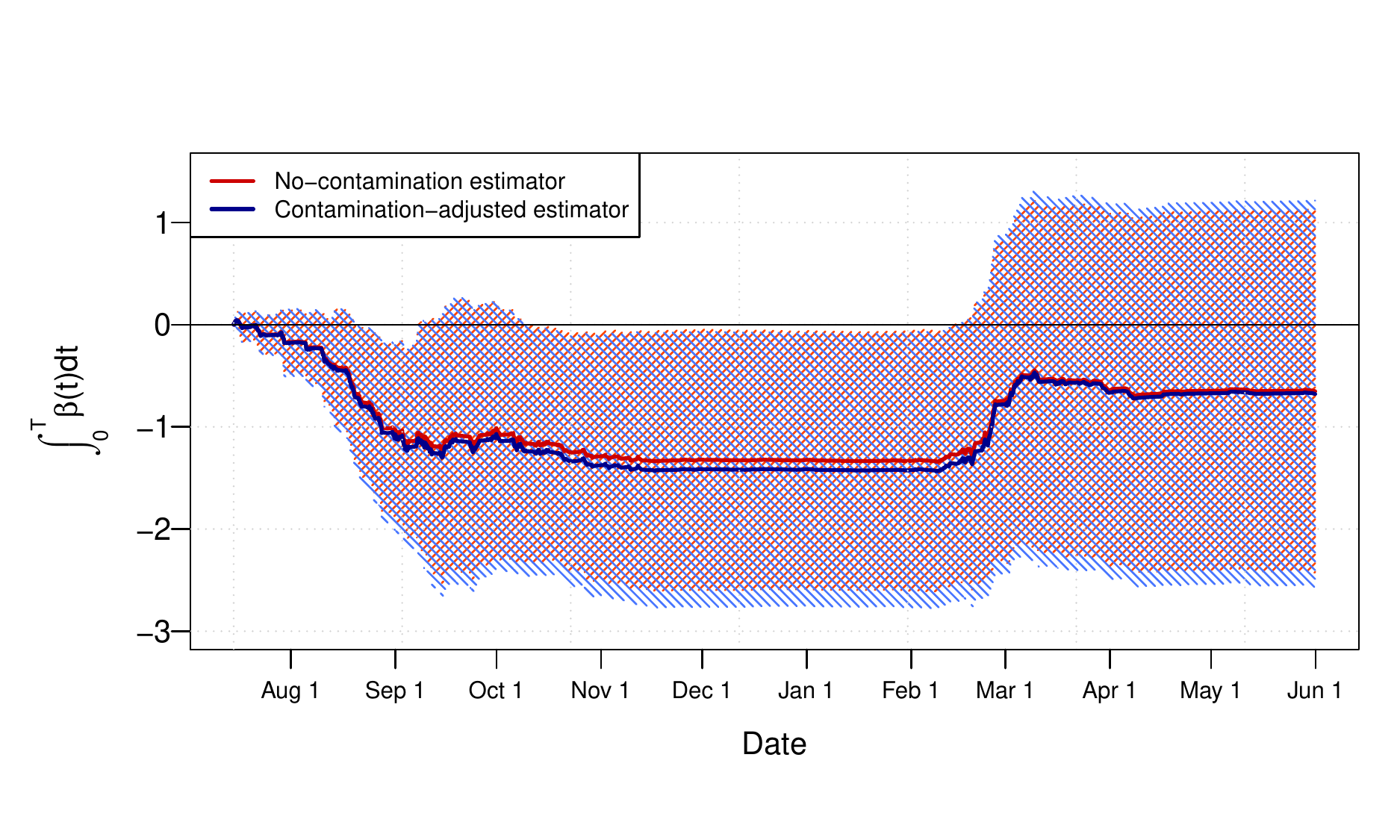}
\end{figure}

\begin{figure}
\caption{\label{fig:infweek}
Weekly observed incidence of lab-confirmed symptomatic influenza infections by type during Year 1 (July 2009 - May 2010) of the study. } 
\centering
\includegraphics[width=7in,height=3.9in]{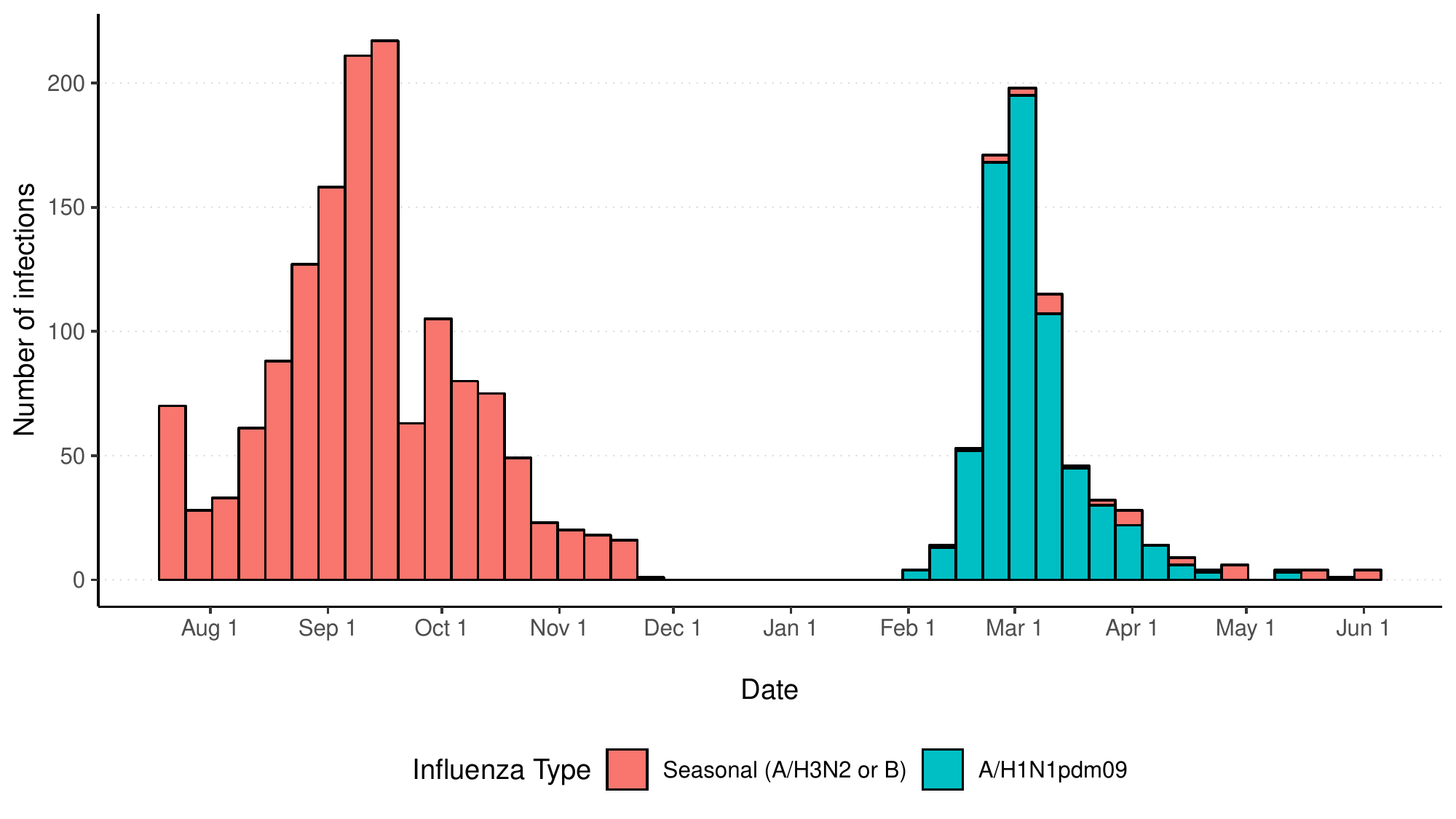}
\end{figure}

Our estimated time-varying treatment effects (both unadjusted and contamination-adjusted) are displayed in Figure~\ref{fig:year1} for Year 1 of the study.  Since the graph displays the integral of the time-varying coefficients, the slopes of the curves represent the coefficients themselves - the estimated difference in hazard rates between vaccine and control villages at each point in time.  Both models indicate that the influenza vaccination program reduces influenza through September.  Then it is estimated to be ineffective until February (since no influenza was circulating), after which the program is associated with an increase in the hazard of influenza for a month.  This latter time period coincides with the appearance of the A (H1N1) (2009) pandemic strain of influenza (A/H1N1pdm09) in the community, which first appeared in late January 2010.  See Figure~\ref{fig:infweek} for a graph of numbers of infections by influenza type and week.

\begin{figure}
\caption{\label{fig:year1_excludingH1N12009}
Estimated effects of the influenza vaccination program for Year 1 (July 2009 - May 2010) of the study, excluding H1N1 2009 pandemic influenza infections. The time-varying effects are the change in the cumulative hazard of lab-confirmed symptomatic influenza infection, measured in percentage points.}
\centering
\includegraphics[width = 1.1\textwidth]{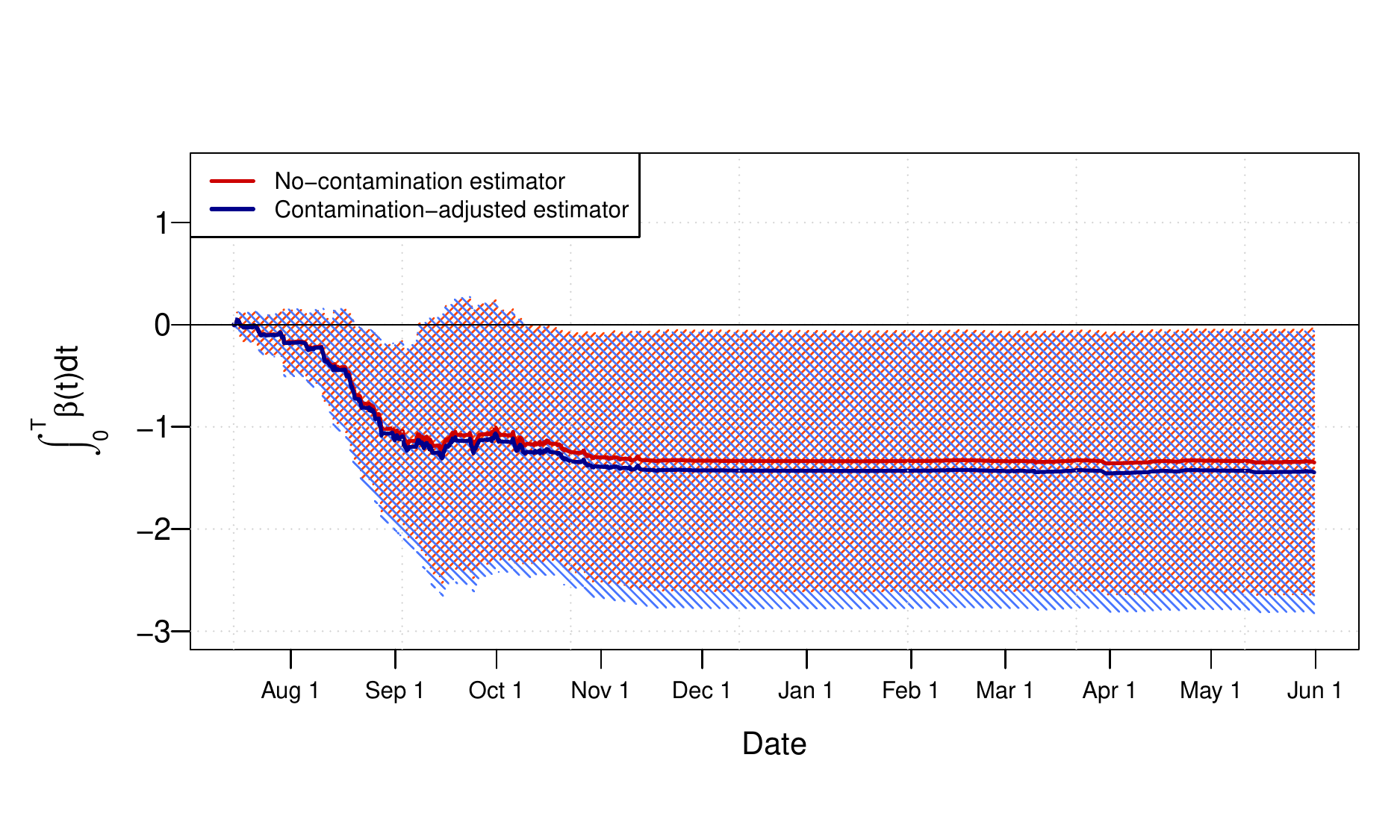}
\end{figure}

Figure~\ref{fig:year1_excludingH1N12009} presents the two estimators excluding cases of A/H1N1pdm09 influenza from the analysis.  Its slope represents the instantaneous effect of the influenza vaccination program on the hazard of infection for non-pandemic strains only. 

\begin{figure}
\caption{\label{fig:year2}
Estimated effects of the influenza vaccination program for Year 2 (July 2010 - May 2011) of the study.  The time-varying effects are the change in the cumulative hazard of lab-confirmed symptomatic influenza infection, measured in percentage points.}
\centering
\includegraphics[width = 1.1\textwidth]{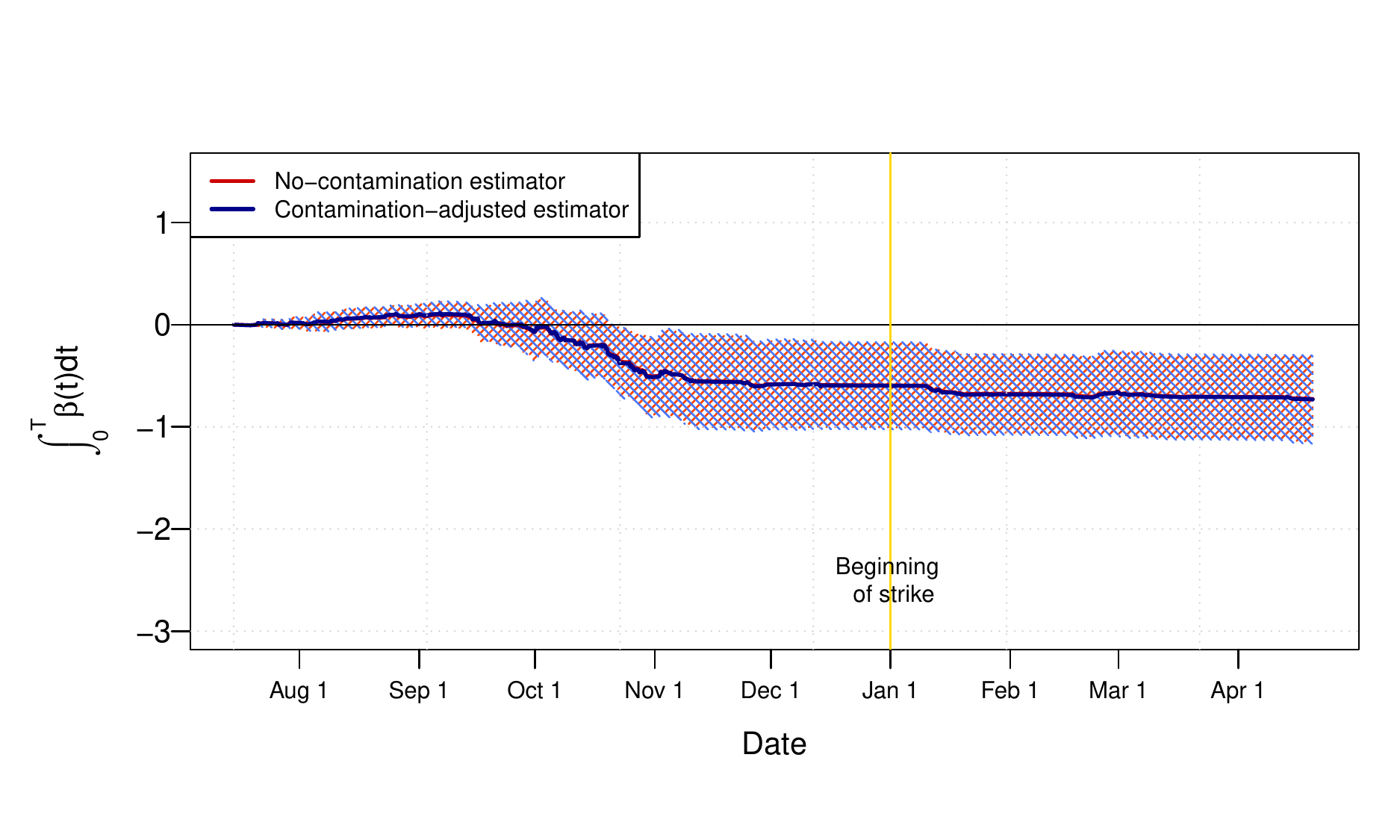} 
\end{figure}

\begin{figure}
\caption{\label{fig:infweek2}
Weekly observed incidence of lab-confirmed symptomatic influenza infections by type during Year 2 (July 2010 - May 2011) of the study. }
\centering
\includegraphics[width=6in,height=3.6in]{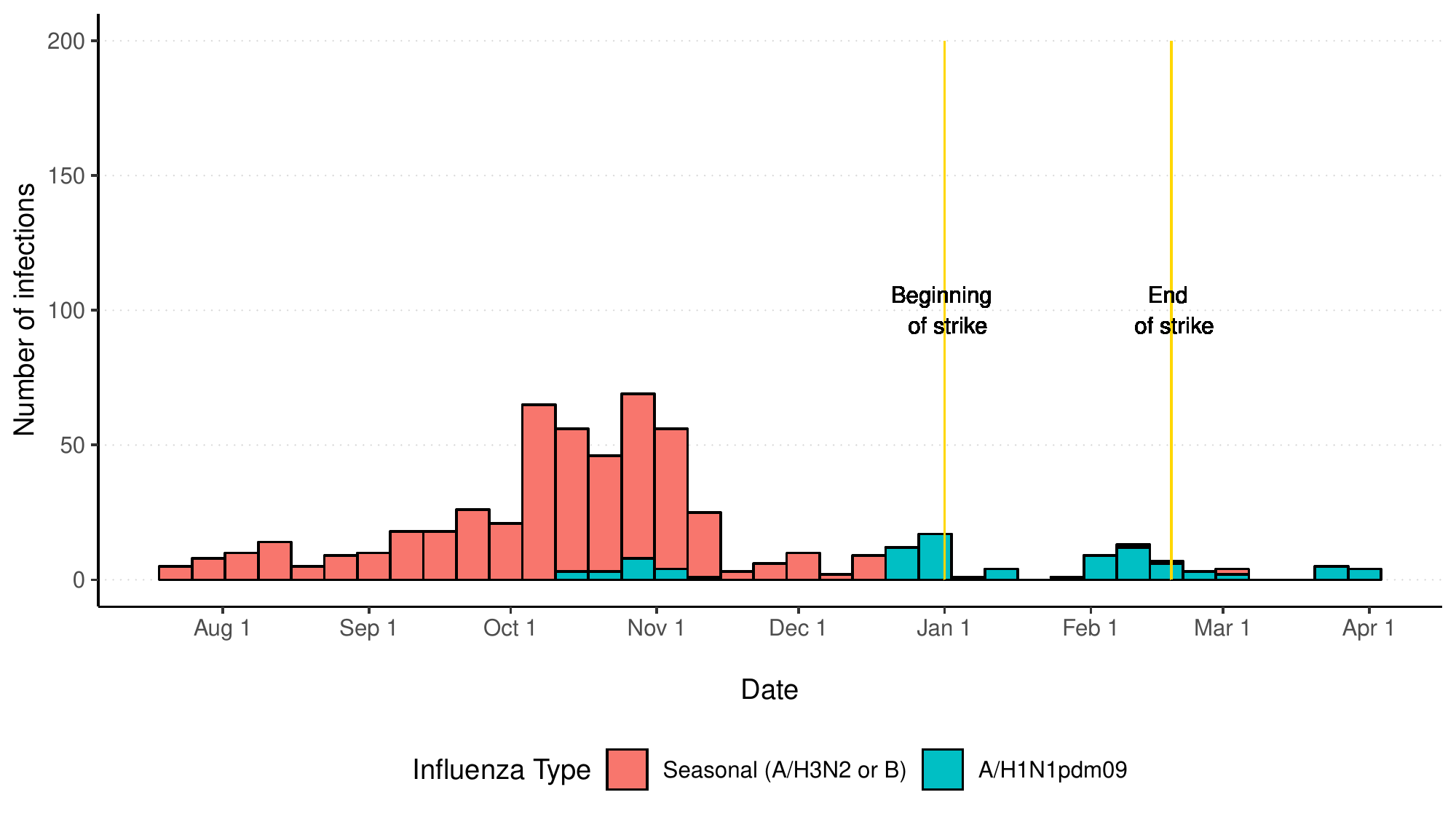}
\end{figure}

Figure~\ref{fig:year2} presents results for the second year of the study.  This is the first publication of Year 2 estimates for this trial, as the primary analysis only analyzed Year 1~\citep{primary}.  As mentioned previously, the formulation of the vaccine provided during this year included the A/H1N1pdm09 strain, unlike the formulation provided in Year 1.  Figure~\ref{fig:infweek2} shows that substantially fewer infections were detected this year.  We expect reports to be lower during the strike (Jan. 1 - Feb. 18, 2011) since household surveillance was not conducted during that time, but frequencies prior to the strike were also much lower than in Year 1.  Figure~\ref{fig:year2} shows that after a delay of approximately two months with little effect, the two estimators both indicate that influenza vaccination reduced incidence in Year 2.  The delay is likely due to the relative sparsity of cases in the first weeks of the year. The start of the strike mentioned in Section~\ref{sec:methods} is shown as a vertical line. 


Table~\ref{tab:cum_inc} displays the estimated difference in cumulative hazard of lab-confirmed symptomatic influenza infection due to the influenza vaccination program.  These are simply the values of the curves in Figures~\ref{fig:year1},~\ref{fig:year1_excludingH1N12009}, and~\ref{fig:year2} for the last day of follow-up, and the confidence intervals correspond to those in the figures.  Because the cumulative hazard is low, the difference in cumulative hazard is approximately equal to the difference in cumulative incidence due to treatment. 

The overall incidence rates are displayed in Table~\ref{tab:fluinc} for comparison purposes.  Since the overall incidence in the control group was 6.13\%, our estimated additive effect of -0.68\% indicates the vaccination program prevented about 11\% of influenza infections.  

 
 Our two estimators and confidence intervals are similar, but the no-contamination estimators are slightly attenuated because they assume no mixing between clusters of opposite treatment assignments.  The confidence intervals for the contamination-adjusted estimator are slightly wider, reflecting the loss of information caused by contamination, but again, are similar.  For Year 1 both effects are not statistically significant when all infections are included but achieve significance (barely) when A/H1N1pdm09 infections are excluded.  The Year 2 estimates are statistically significant.  The Year 2 estimates are interpreted differently as they cover different time intervals; a higher difference in cumulative incidence is expected for the longer interval if vaccine performance stays the same.  While bias from the strike starting Jan 1, 2011 does not impact the Year 2 estimate censored at that date, the uncensored one could be biased. The rates of reporting infections during health post visits (as opposed to household visits) were 12.5\% in treated villages and 17\% in control villages, so the vaccine effect could be overestimated by including a time interval with only health post visits.  Because the rates are similar, and because the strike lasted 49 days of a 320-day follow-up period, the bias is likely low.
\begin{table}
\caption{\label{tab:fluinc}Incidence of influenza by treatment group and study year.}
\centering
\begin{tabular}{lccc}
  \hline
Study Year & Treated & Control & All \\ 
  \hline
Year 1, all infections & 999/18200 (5.49\%) & 1076/17550 (6.13\%) & 2075/35750 (5.8\%) \\ 
  Year 1, excluding A/H1N1pdm09 & 630/18200 (3.46\%) & 833/17550 (4.75\%) & 1463/35750 (4.09\%) \\ 
  Year 2, all infections & 224/18547 (1.21\%) & 341/17815 (1.91\%) & 565/36362 (1.55\%) \\ 
  \hline
\end{tabular}
\end{table}

\begin{table}
\caption{\label{tab:cum_inc}
Estimated difference in cumulative incidence of influenza (measured in percentage points) due to the influenza vaccination program.}
\centering
\begin{tabular}{lccccc}
  \hline
   \multicolumn{1}{c}{ }&  \multicolumn{2}{c}{Contamination-Adjusted}&  \multicolumn{2}{c}{No-Contamination}\\  
Study Year & Estimate & 95\% C.I. & Estimate & 95\% C.I. \\ 
  \hline
  Year 1, all infections & -0.68 & [-2.53, 1.18] & -0.65 & [-2.40, 1.09] \\ 
  Year 1, excluding A/H1N1pdm09 & -1.45 & [-2.81, -0.08] & -1.35 & [-2.64, -0.06] \\ 
   Year 2 (July - Dec 2010) & -0.59 & [-1.01, -0.17] & -0.59 & [-0.99, -0.19] \\ 
  Year 2 (July 2010 - May 2011) & -0.73 & [-1.16, -0.31] & -0.73 & [-1.14, -0.32] \\
\hline
\end{tabular}
\end{table}

\section{Discussion}
\label{sec:concl}

 We have applied state-of-the art statistical methodology to estimate the overall effect of a trivalent influenza vaccine program in Niakhar, Senegal.  This method incorporates social contact data together with treatment and infection data to reduce the bias in this estimate caused by interference between clusters.   Ours is the first study we know of applying this novel method to contact and infection data collected jointly in a clinical trial setting.  We produce the first estimates of contact rates between clusters of opposite treatment assignments for this trial and the first, to our knowledge, in Senegal.  Our results provide insight into the extent to which the standard assumption of partial interference is violated in a trial of this structure and of the impact of this violation on estimates.  
 
 Our time-varying effect estimates show that in Year 1 of the study, the treatment program -- vaccination of children -- reduced lab-confirmed symptomatic infection with seasonal influenza in the community.  Our estimates found the treatment program to be associated with a small (though statistically insignificant) increase in infections with A/H1N1pdm09 influenza.  While other studies have found evidence for this relationship~\citep{cowling2009, skowronski2010association}, others have found evidence that trivalent influenza vaccination protects against A/H1N1pdm09 infection.  A meta-analysis of 17 studies, including the two just mentioned, found that the overall evidence points to a protective effect, but the authors cautioned against drawing a solid conclusion because most of the studies reviewed were observational~\citep{yin2012impacts}.  Two subsequent randomized trials also found evidence for a protective effect~\citep{cowling2012protective, mcbride2016efficacy}.

The extent of contamination measured in our data resulted in little difference between the cumulative incidence for the estimator adjusting for contamination and the one assuming no contamination. The latter was smaller because, as has been found in other studies, contacts to members of clusters of the opposite assignment attenuate the estimate of the overall effect from what it would have been with no contamination~\citep{carnegie2016estimation, TionoOuedraogo2013, WangGoyal2014}.
The model we implement explicitly adjusts for contamination, correcting this under-estimation.  In addition, the standard errors associated with this adjusted estimator were larger than those for the no-contamination estimator because information available to estimate the effect of the treatment program decreases as mixing increases -- so these intervals accurately reflect the  decrease in information from zero mixing to the small level of mixing we observed. As noted in~\cite{carnegie2016estimation} the approach we have used to estimate the overall effect fails when 50\% of contacts occur to clusters of the opposite treatment assignment.  This is because our method uses the contact rates between clusters to differentiate treatment status, so no information distinguishing clusters is available for our approach when mixing is at 50\%.  

The level of contamination in the data was fairly small: the percent of contacts to clusters of the opposite treatment was between 0\% and 3\% for most villages, although there were some outliers, with 14\% being the largest observed value.  To our knowledge, these are the first data-based contamination estimates of this type for  Senegal.  Our finding that this amount of contamination has a negligible impact on the effect estimate may be encouraging for researchers who carefully define cluster selection to minimize contamination, as was done in this study.  The villages in this trial were separated by physical boundaries such as bodies of water and roads, and their definition as cultural/political entities also has an impact on social contact behavior.  
 
Our study has several limitations.  First, the extent of missing data in the contact survey is substantial.  As noted previously, for locations visited outside the home two days before the survey, 24\% are missing time of day, 59\% are missing the number of people contacted, and 32\% do not have a village number recorded.  We used multiple imputation to adjust for missing data, but bias is still a risk.  For example, if numbers of people contacted in villages of the opposite assignment were higher for participants who did not respond to this question than for those who responded (and who have similar values for covariates included the multiple imputation model), then the true contamination values may be higher than our predicted values.  This would mean that the magnitude of the true overall effect is larger than our estimate.  If, on the other hand, we have overestimated contamination, then the true effect may be closer to our no-contamination estimate (closer to -0.65 than -0.68). 
Implementation of similar surveys in the future may be improved by a diary-based approach, in which participants fill out a paper diary as they go about their day~\citep{polymod, beraud2015french, melegaro2017social, johnstone2011social, horby, fu2012representative, read2014social}. In addition we would recommend consideration of procedures employed by~\cite{kiti2014quantifying}, including conducting a pilot study, providing wristwatches with pre-programmed alarms to remind participants to fill out their diary, and by assigning ``shadow'' respondents to fill out the diary for illiterate participants.  Alternately and potentially more accurate would be an approach using remote wireless sensors to detect when two  participants are located within 1.5 meters of each other - a distance at which infection may be transmitted~\citep{kiti2016quantifying, stehle, fournet2014contact, barclay2014positive, genois2015data}.

A second limitation of the contact survey is that contacts were reported separately for morning and afternoon time intervals without recording the extent of overlap.  Because morning and afternoon contamination estimates were similar, either is likely a reasonable approximation to the percent of contacts to clusters of the opposite assignment during a full day.  However, it would be preferable to record numbers of contacts throughout the entire day in future studies. We also note that contacts recorded on the day of the survey did not contribute to analysis since truncation bias arose from the fact that  most surveys were conducted in the morning.  A diary-based approach would avoid this problem, or if interviews are conducted, they should focus on days before the survey day.  The literacy level of the population of interest should be considered in choosing the optimal approach to collect contact data.  

Finally, the type of contacts recorded in our study emphasize transmission via large droplets (in close proximity) rather than by aerosol droplets which have a longer range.  While many studies have investigated the importance of fomite transmission, physical contacts, small droplets, and aerosol droplets for transmission, their relative importance is not well understood~\citep{weber2008inactivation, cowling2013aerosol,teunis2010high,wei2016airborne,kutter2018transmission}.
Although the contact survey had limitations, it seems unlikely that the true contamination levels are higher enough than our estimated ones to substantially impact the efficacy estimates. 
Therefore we believe that our conclusion that contamination was low and had only a small impact on efficacy estimates is valid.  However, careful design of the contact survey would improve data precision if a similar approach is applied when clusters are smaller and closer.  We would recommend such studies as future research.  For example, a compound-based randomization scheme had been considered for this trial design instead of village-based, and in fact, the protocol allowed for both possibilities.  The level of contamination for such a design, which would likely be higher than that for villages, could be estimated with our social network data in order to understand its potential impact on estimation.  Although our method adjusts for the contamination, higher contamination decreases the information available to detect an effect.  Since our approach removes the dilution from the effect estimate while simultaneously increasing standard errors, the lost power from contamination is not regained via our adjustment.  Rather, the estimate and standard error estimates are both more accurate than unadjusted estimates. We expect this relationship to hold for other adjustment approaches which have been proposed but, to our knowledge, not yet applied or tested (e.g.,~\cite{reiner2016quantifying}).

We also recommend collection and estimation of cross-cluster contamination for different types of contacts (e.g., physical contacts, sexual contacts), for various definitions of clusters in various settings.  These estimates can be used to inform future trial designs, choose whether the method we have applied would be better than one which does not adjust for contamination, and ultimately improve the accuracy of vaccine effectiveness and standard error estimates.




 \bibliographystyle{elsarticle-harv}
\bibliography{causal_interference,epi_models,survival_analysis}

\appendix

\newgeometry{left=1cm,bottom=3cm,top=2cm,right=.5cm}
\begin{landscape}

\section{Contact survey form}
\begin{figure}[h]
    \centering
\includegraphics[width=8
in, height=5.5in]{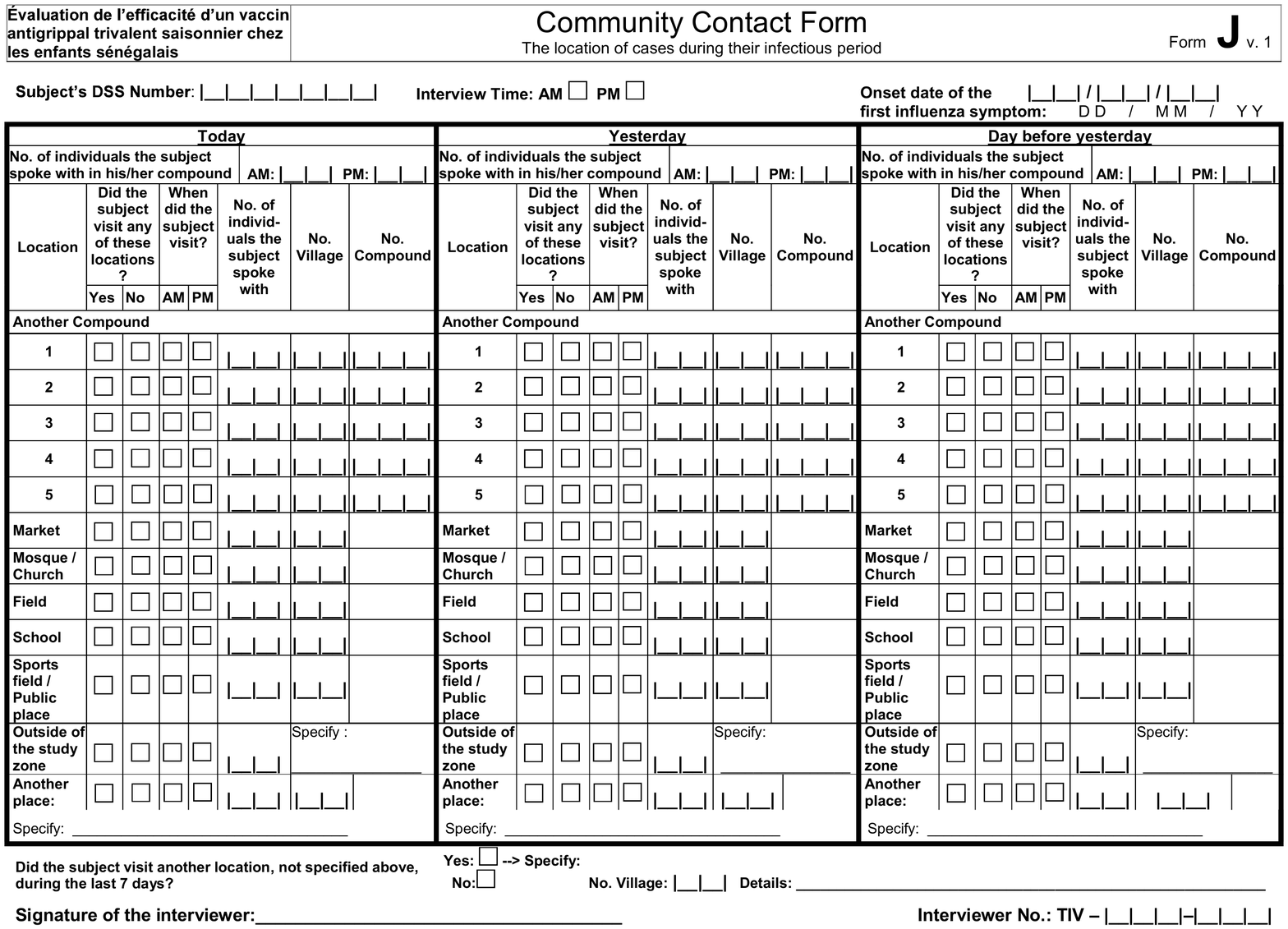} 
\end{figure}
\end{landscape}
\restoregeometry{}

\section{Cross-village exposure summaries}
\label{appendix-sec1}

This section displays analyses that informed our calculation of cross-village exposure rates. The cross-village exposure rate for a village is defined to be the percentage of contacts to people in clusters of the opposite treatment assignment. The tables in this section summarize these rates based on contacts made while the respondent was visiting other villages and do not incorporate contacts made to visitors from other villages in the respondent's own home.  The tables summarize rates of contacts to treated villages by village of residence; these represent the cross-village exposure rate for control villages and one minus the cross-village exposure rate for treated villages.

Table~\ref{tab:am_pm} compares fractions and percentages of contacts to treated villages between morning and afternoon/evening time intervals.  Cross-cluster exposure rates are similar for the two time intervals, with the main differences being Poudaye and Logdir, whose higher variability than others is likely due to the small number of overall contacts reported in those villages.

Table~\ref{tab:symp_asymp} compares fractions and percentages of contacts reported by respondents during visits to treated villages during the morning by village number and symptom status.  Since numbers of asymptomatic reports are low and cross-village exposure is low, cross-village exposure is lower for asymptomatic than symptomatic participants in most villages.  Table~\ref{tab:symp_asymp_imp} shows the analogous percentages calculated based on the imputed data and shows higher levels of cross-cluster exposure for symptomatic than asymptomatic people.

 \begin{table}
  \caption{\label{tab:am_pm}Fraction and percent of contacts reported by respondents while located in treated villages by village of residence and time of day. }
 \centering
 \begin{tabular}{lccccc}
   \hline
 Village & Treatment & AM Fraction & AM Percent & PM Fraction & PM Percent \\ 
   \hline
Darou & Vaccine & 149/149 & 100 & 123/123 & 100 \\ 
  Kalome Ndofane & Vaccine & 1221/1244 & 98 & 959/959 & 100 \\ 
  Poudaye & Vaccine & 53/65 & 82 & 47/47 & 100 \\ 
  Mokane Ngouye & Vaccine & 1478/1513 & 98 & 1307/1312 & 100 \\ 
  Ngayokheme & Vaccine & 6414/6489 & 99 & 5638/5682 & 99 \\ 
  Ndokh & Vaccine & 135/137 & 99 & 101/103 & 98 \\ 
  Nghonine & Vaccine & 303/306 & 99 & 222/231 & 96 \\ 
  Ngangarlame & Vaccine & 859/953 & 90 & 533/586 & 91 \\ 
  Diohine & Vaccine & 1195/1377 & 87 & 1104/1230 & 90 \\ 
  Logdir & Vaccine & 158/185 & 85 & 53/80 & 66 \\ \hline
  Ngalagne Kop & Control & 0/1052 & 0 & 0/873 & 0 \\ 
  Bary Ndondol & Control & 0/545 & 0 & 0/512 & 0 \\ 
  Mboyene & Control & 1/634 & 0 & 1/522 & 0 \\ 
  Toucar & Control & 6/2870 & 0 & 0/2006 & 0 \\ 
  Godel & Control & 0/456 & 0 & 0/449 & 0 \\ 
  Khassous & Control & 0/151 & 0 & 0/96 & 0 \\ 
  Kothioh & Control & 3/643 & 0 & 0/440 & 0 \\ 
  Meme & Control & 0/78 & 0 & 0/97 & 0 \\ 
  Poultok Diohine & Control & 0/1717 & 0 & 0/1240 & 0 \\ 
  Gadiak & Control & 10/466 & 2 & 10/310 & 3 \\     
  \hline
 \end{tabular}
\end{table}

\begin{table}
\caption{\label{tab:symp_asymp}Fraction and percent of contacts reported by respondents while located in treated villages by village of residence and symptom status. }
\centering
\begin{tabular}{lccccc}
  \hline
 \multicolumn{1}{l}{ }&  \multicolumn{1}{c}{ Treatment}&  \multicolumn{2}{c}{Asymptomatic} &  \multicolumn{2}{c}{Symptomatic}\\
Village & Assignment & Fraction & Percent & Fraction & Percent \\ 
  \hline
Darou & Vaccine & 61/61 & 100 & 88/88 & 100 \\ 
  Ndokh & Vaccine & 28/30 & 93 & 107/107 & 100 \\
  Ngayokheme & Vaccine & 1468/1490 & 99 & 4946/4999 & 99 \\ 
  Nghonine & Vaccine & 32/32 & 100 & 271/274 & 99 \\ 
  Kalome Ndofane & Vaccine & 352/359 & 98 & 869/885 & 98 \\ 
  Mokane Ngouye & Vaccine & 178/183 & 97 & 1300/1330 & 98 \\ 
  Ngangarlame & Vaccine & 171/174 & 98 & 688/779 & 88 \\ 
  Diohine & Vaccine & 555/610 & 91 & 640/767 & 83 \\ 
  Poudaye & Vaccine & 6/6 & 100 & 47/59 & 80 \\ 
  Logdir & Vaccine & 58/58 & 100 & 100/127 & 79 \\ 
  \hline
  Ngalagne Kop & Control & 0/258 & 0 & 0/794 & 0 \\ 
  Bary Ndondol & Control & 0/4 & 0 & 0/541 & 0 \\
  Mboyene & Control & 0/86 & 0 & 1/548 & 0 \\ 
  Toucar & Control & 0/839 & 0 & 6/2031 & 0 \\
  Godel & Control & 0/316 & 0 & 0/140 & 0 \\ 
  Khassous & Control & 0/19 & 0 & 0/132 & 0 \\
  Meme & Control & 0/30 & 0 & 0/48 & 0 \\ 
  Poultok Diohine & Control & 0/559 & 0 & 0/1158 & 0 \\ 
  Kothioh & Control & 0/157 & 0 & 3/486 & 1 \\
  Gadiak & Control & 0/137 & 0 & 10/329 & 3 \\
 \hline
\end{tabular}
\end{table}

\begin{table}
\caption{\label{tab:symp_asymp_imp}
Percent of contacts reported by respondents while located in treated villages by village of residence and symptom status based on multiply imputed data. }
\centering
\begin{tabular}{lcccc}
  \hline
  \multicolumn{1}{l}{ }&  \multicolumn{1}{c}{ Treatment}&  \multicolumn{2}{c}{} \\
Village &  Assignment & All & Asymptomatic & Symptomatic \\ 
  \hline
Kalome Ndofane & Vaccine & 100 & 100 & 100 \\
Ngangarlame & Vaccine & 99 & 100 & 99 \\ 
Diohine & Vaccine & 99 & 100 & 98 \\ 
Mokane Ngouye & Vaccine & 99 & 100 & 99 \\ 
Ngayokheme & Vaccine & 99 & 99 & 99 \\ 
Ndokh & Vaccine & 99 & 99 & 99 \\ 
Nghonine & Vaccine & 98 & 99 & 98 \\ 
Logdir & Vaccine & 95 & 94 & 96 \\ 
Darou & Vaccine & 96 & 90 & 100 \\ 
Poudaye & Vaccine & 93 & 84 & 95 \\ 
  \hline
Ngalagne Kop & Control & 0 & 0 & 0 \\ 
Bary Ndondol & Control & 0 & 0 & 0 \\ 
Mboyene & Control & 0 & 0 & 0 \\ 
Poultok Diohine & Control & 0 & 1 & 0 \\ 
Toucar & Control & 1 & 1 & 0 \\ 
Gadiak & Control & 2 & 3 & 2 \\ 
Godel & Control & 2 & 1 & 3 \\ 
Khassous & Control & 3 & 0 & 3 \\ 
Kothioh & Control & 3 & 2 & 4 \\ 
Meme & Control & 14 & 0 & 20 \\     
  \hline
   \hline
\end{tabular}
\end{table}

\clearpage

\section{Calculation of time-to-event}
\label{subsection:tte}

We restrict our analysis to the twenty villages enrolled in the cluster-randomized trial as these villages received both active and passive surveillance while the other ten received only passive surveillance.  The surveillance period for Year 1 was July 15, 2009 to May 31, 2010.  These dates determined the start and end of follow-up participants with the following exceptions:

\begin{itemize}
	\item Start of follow-up was the date participants moved to the study area if the move took place after surveillance began.
	\item If participants moved out of the study area or to a cluster of the opposite treatment assignment during surveillance, their end of follow-up was the move date.
\end{itemize}

Time-to-infection was calculated by subtracting the start of follow-up from the sample collection date for infected people; censoring times were calculated based on start and end of follow-up for uninfected people.

Thirteen participants were excluded from analysis because of inconsistencies in their recorded residence data.  In addition, those who moved to the study area after the end of Year 1, and those who were infected before moving to the study area or before follow-up began were excluded.  Because the primary analysis did not censor or exclude participants based on their residence data, our counts of participants and cases differ slightly from that paper~\citep{primary}.


Time to event for Year 2 (for which surveillance covered July 15, 2010 to May 31, 2011) was calculated analogously.  However, during Year 2 of the study, household-based surveillance did not occur from January 1, 2011 to February 18, 2011 due to a strike of employees performing this surveillance, so only infections reported in health posts were recorded during that time period. This could cause bias if the proportions of infections observed at home compared to in health posts different between treatment arms.  During the non-strike period of Year 2, proportions of lab-confirmed symptomatic influenza infections reported during household visits were $83.07\%$ in the control group and $87.50\%$ in the vaccine group, respectively (Table~\ref{tab:differential}).  Since infections for control arm participants were reported more frequently in health posts than those for vaccine arm participants, the differential reporting could create bias in the efficacy estimate, making the vaccine appear more effective than it actually is.  Inverse probability weighting was considered to correct this bias~\citep{seaman2013review}.  Such an approach would entail up-weighting the observed infections during the strike by $\frac{1}{0.1693}=5.91$ in the control arm and $\frac{1}{0.1250}=8.00$ in the vaccine arm, and down-weighting the people classified as uninfected throughout the study period (since some of these would have had infections that would have been detected during household surveillance during the strike).  This approach would assume that health post visiting behavior was the same during the strike and outside of the strike.  However, the data indicate that that assumption does not hold.  Outside of the strike, 66\% of infections reported in health posts were in the control group, but during the strike, 78\% were.  The larger proportion of up-weighted control group infections resulted in a weighted overall effect estimate that was higher, rather than lower, than the unweighted one.  As the assumption required by inverse probability weighting did not hold, we instead censored the Year 2 data at the last day before the strike.  A secondary analysis includes all of the Year 2 data. 



\begin{table}
\caption{Reporting rates of lab-confirmed symptomatic infection by location within each treatment arm during Year 2, excluding the strike period \label{tab:differential}}
\centering
\begin{tabular}{lcc}
\hline
 & Control & Vaccine \\ 
  \hline
Percent reported in compounds & 83.07 & 87.50 \\ 
  Percent reported in health posts & 16.93 & 12.50 \\   \hline
\end{tabular}
\end{table}

\section{Correspondence to compartmental model for infectious disease transmission}

The additive hazards model applied in this paper has a natural correspondence to an SIR (Susceptible-Infected-Removed) compartmental model for disease transmission.  To see this, recall that the contamination-adjusted estimator for an individual in cluster $j$ is obtained from the following additive hazards model:
\[
\lambda_j(t|M) = \beta_0(t) + \beta_M(t)m_j,
\]
where $m_j$ is the total percentage of contacts of susceptibles in cluster $j$ that are with treated clusters.  

Next, we define the following notation:
\begin{enumerate}
    \item $Y_k(t)$ = the number of infected people in cluster $k$ at time $t$
    \item $\kappa$ = the overall average contact rate
    \item $\eta_k$ = the per-contact transmission probability of infectives in cluster $k$
    \item $m_{jk}$ = the percentage of contacts from people in cluster $j$ with those in  cluster $k$
    \item $\alpha_{jk}$ = the rate of new infections among susceptibles in cluster $j$ from infectives in cluster $k$
    \item  $N_k$ = the population size of cluster $k$. For simplicity, we assume a fixed population size in each cluster. 
\end{enumerate}  
The SIR compartmental model assumes that the rate of transmission from infectives in cluster $k$ to susceptibles in cluster $j$ is the product of the overall contact rate, the percentage of contacts from cluster $j$ that are to people in cluster $k$, and the per-contact
transmission probability:  $\alpha_{jk} = \kappa m_{jk}\eta_k$.  The hazard function of a susceptible in cluster $j$ is found by summing these cluster-specific transmission rates, weighted by their cluster-specific proportions of infectives, across all clusters:
\begin{equation}
\lambda_j(t) = \sum_{k=1}^c \alpha_{jk} \frac{Y_k(t)}{N_k}  \label{eq:haz1}    
\end{equation}

To simplify notation, we define $\nu_j(t) = \kappa \eta_j \frac{Y_j(t)}{N_j}$, so

\begin{equation}
\lambda_j(t) = 
\sum_{k=1}^c \alpha_{jk} \frac{Y_k(t)}{N_k} =
\sum_{k=1}^c m_{jk}\nu_k(t) \label{eq:haz}    
\end{equation}

The estimand of interest, which we will denote $\beta(t)$, is the population-averaged difference in hazard of infection associated with a change from 0\% to 100\% exposure to treatment.  That is, $\beta(t) = \bar{\nu}^{trt}(t)- \bar{\nu}^{ctr}(t)$, where $\bar{\nu}^{trt}(t)$ is the average of $\nu(t)$ in treated clusters and  $\bar{\nu}^{ctr}(t)$ is the average of $\nu(t)$ in control clusters.  While $\hat{\beta}_Z(t)$ is a consistent estimator for $\beta(t)$ in the absence of contamination, Carnegie, Rui, and Wang proved that $\hat{\beta}_M(t)$ is a consistent estimator for $\beta(t)$ in the presence of measured  contamination.~\citep{carnegie2016estimation}  

If we assume that the individual hazards of infection are identical within a cluster, then the instantaneous rate of change of the number of infected individuals in cluster $i$ at time $t$ is found by summing the individual hazards of all susceptibles in cluster $i$. Letting $S_i(t)$ denote the number of susceptibles in cluster $i$ at time $t$ and substituting from \eqref{eq:haz} yields:
$$\frac{d Y_i(t)}{dt} =  S_i  \sum_{j=1}^c \alpha_{ij} \frac{Y_j(t)}{N_j} =   \sum_{j=1}^c \alpha_{ij} \frac{Y_j(t)S_i(t) }{N_j},$$
which corresponds to an SIR model with no birth and or death.  A similar expression for the rate of change of susceptibles is analogously derived, and generalizations such as birth, death, and the addition of an exposed state for an SEIR model are addressed in~\cite{carnegie2016estimation}. 

\newpage
\section{Rationale for adjustment in estimated contamination estimates based on reports from visitors to the respondent's compound}

\begin{figure}
\caption{\label{fig:toynetwork}Toy example of a social network with corresponding adjacency matrices.}
\centering
\includegraphics[width=4.5in, height=7in]{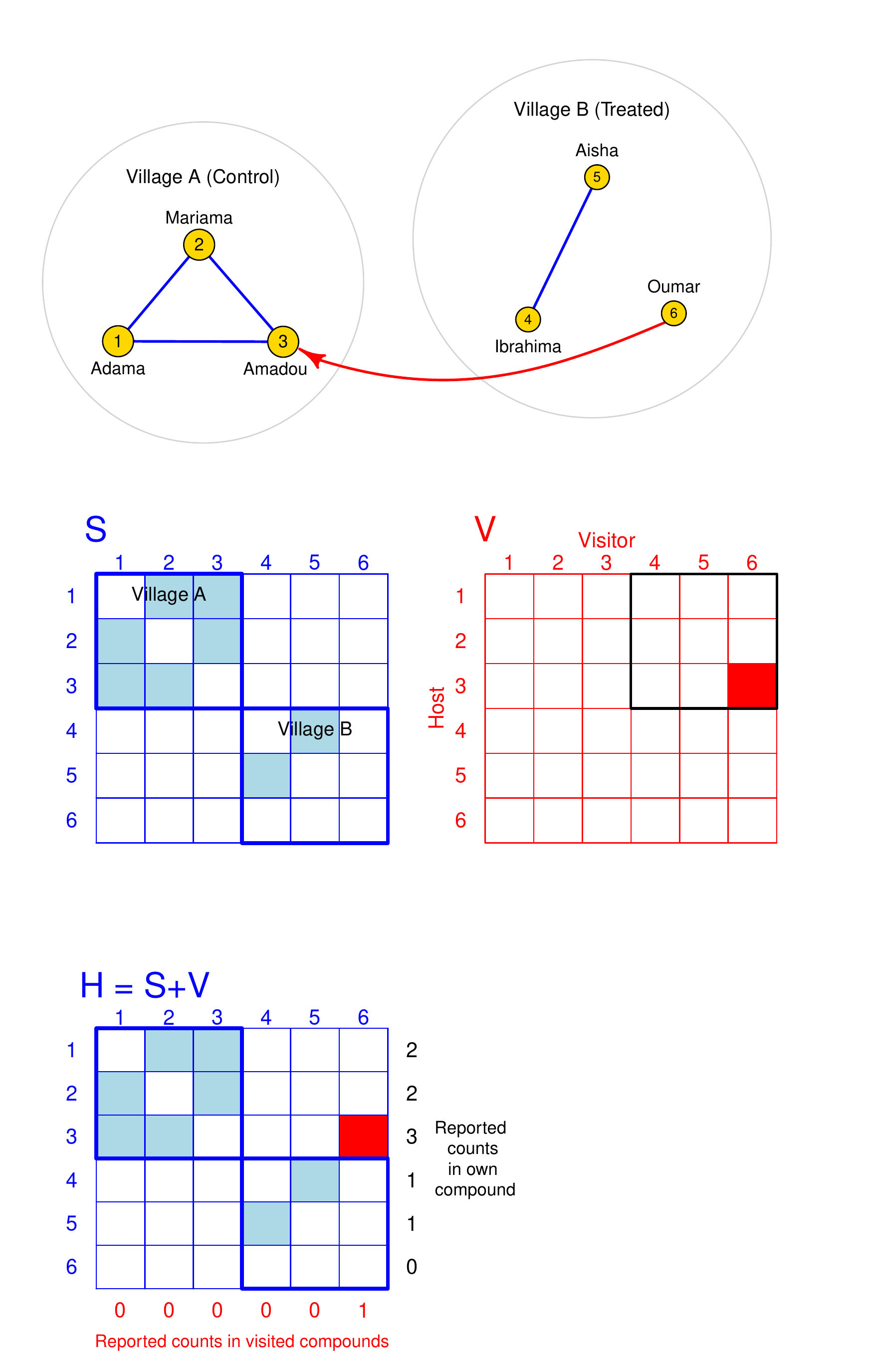}
\end{figure}

We define the following notation as described in the main text:
\begin{itemize}
    \item $n_j = $ number of people living in cluster $j$
    \item $D_i =$ number of contacts reported by person $i$ 
    \item $T_i =$ number of contacts person $i$ made in a location in a treated cluster. 
    \item $p_j=$ the proportion of contacts from cluster $j$ to treated clusters.
    \item $V_{trt,j} = $ the total number of contacts reported by people in any treated cluster during their visits to compounds in cluster $j$. 
\end{itemize}
We initially estimated $p_j$ with $$\hat{p_j}=\frac{\sum_{i=1}^{n_j}T_i }{\sum_{i=1}^{n_j}D_i}$$   
The numerator does not include contacts occurring within the respondent's own compound to visitors from other clusters, since these occurred within the respondent's assigned cluster. We can use estimates reported by visitors from clusters of the opposite assignment, rather than by respondents in cluster $j$, to obtain this information. When $j$ is a control cluster, our estimator is appropriately updated by adding the percent of contacts from treated clusters to compounds in cluster $j$ to the contamination estimate:
$$\hat{p_j}=\frac{\sum_{i=1}^{n_j}T_i +V_{trt,j}}{\sum_{i=1}^{n_j}D_i} = \frac{\sum_{i=1}^{n_j}T_i}{\sum_{i=1}^{n_j}D_i}+\frac{V_{trt,j}}{\sum_{i=1}^{n_j}D_i}$$
To understand this, we will walk the reader through a toy example of a network depicted in Figure~\ref{fig:toynetwork}, a diagram similar to that in~\citep{potter2019networks}.  

Here, A is a control village and B is a treated village, and the red arrow indicates that Oumar contacted Amadou while visiting Amadou's home in village A.  For simplicity, assume all network members are surveyed.  The true cross-cluster exposure value for $A$ is 1/7 (noting that each within-village contact is reported twice); it is  $1/3$ for B.  Also for simplicity, our example omits contacts occurring in non-home locations (e.g. market, mosque, etc.), as the proposed adjustment to our estimator does not change how these contribute to the estimates.

The depicted network is not completely observed since respondents did not report the identity of their contacts.  We will define adjacency matrices to illustrate how the completely observed network relates to the recorded data.  Define $S$ to be an adjacency matrix indicating contacts to members of one’s own village, so $S_{ij}=1$ if $i$ and $j$ made contact and belong to the same village.  $S$ is symmetric, since if $i$ contacted $j$, then $j$ contacted $i$ as well.   Let $V$ denote contacts reported while a member of one cluster was visiting a member of a cluster in the opposite treatment arm in the latter’s compound.  $V$ is asymmetric to distinguish the host from the visitor and to align with the way these contacts were reported, and $V_{3,6}=1$ since person 6 visited person 3 in the home of person 3.  The recorded counts of contacts occurring in the respondent's own compound are the row sums of $H \equiv S+V$.  The recorded counts of contacts while the respondent was visiting compounds in villages of the opposite treatment assignment are the column sums of $V$.  Our preliminary approach to estimating interference (without the proposed adjustment) would calculate as follows:

\begin{itemize}

\item For village A, the denominator (total number of contacts) is the sum of the row sums of rows 1, 2, 3 of $H$ (total number of contacts reported in the respondent's home) plus the column sums of columns 1, 2, and 3 of $V$ (total number of contacts while the respondent visited a home in a cluster of the opposite assignment), so the denominator is 7.  The numerator is the sum of column sums of columns 1, 2, 3 of $V$, so the numerator is zero.  Our cross-cluster exposure estimate is 0/7.

\item An analogous approach for village B yields a cross-cluster exposure estimate of 1/3.  

\end{itemize}
Our cross-cluster exposure estimate for A is incorrect since it does not account for the contamination while Oumar was visiting Amadou since it occurred in Amadou's home.   

Our proposed adjustment is to subtract from the numerator of A contacts reported by members of B and occurring in compounds within cluster A.  These comprise the upper right quadrant of matrix V, shown in black, whose sum is 1.  Thus our adjusted estimate for cross-cluster contamination for cluster A is 1/7, the correct value.  A similar adjustment for B involves the lower left quadrant of V; whose sum is zero, so the estimate for B (which was already accurate) remains the same.

\begin{figure}
\caption{\label{fig:locsymp}Location distribution of contacts by symptom status and time interval.  Note: This figure has been published in~\citep{potter2019networks} and is reproduced with permission of the authors.}
\centering
\includegraphics[width = 0.85\textwidth]{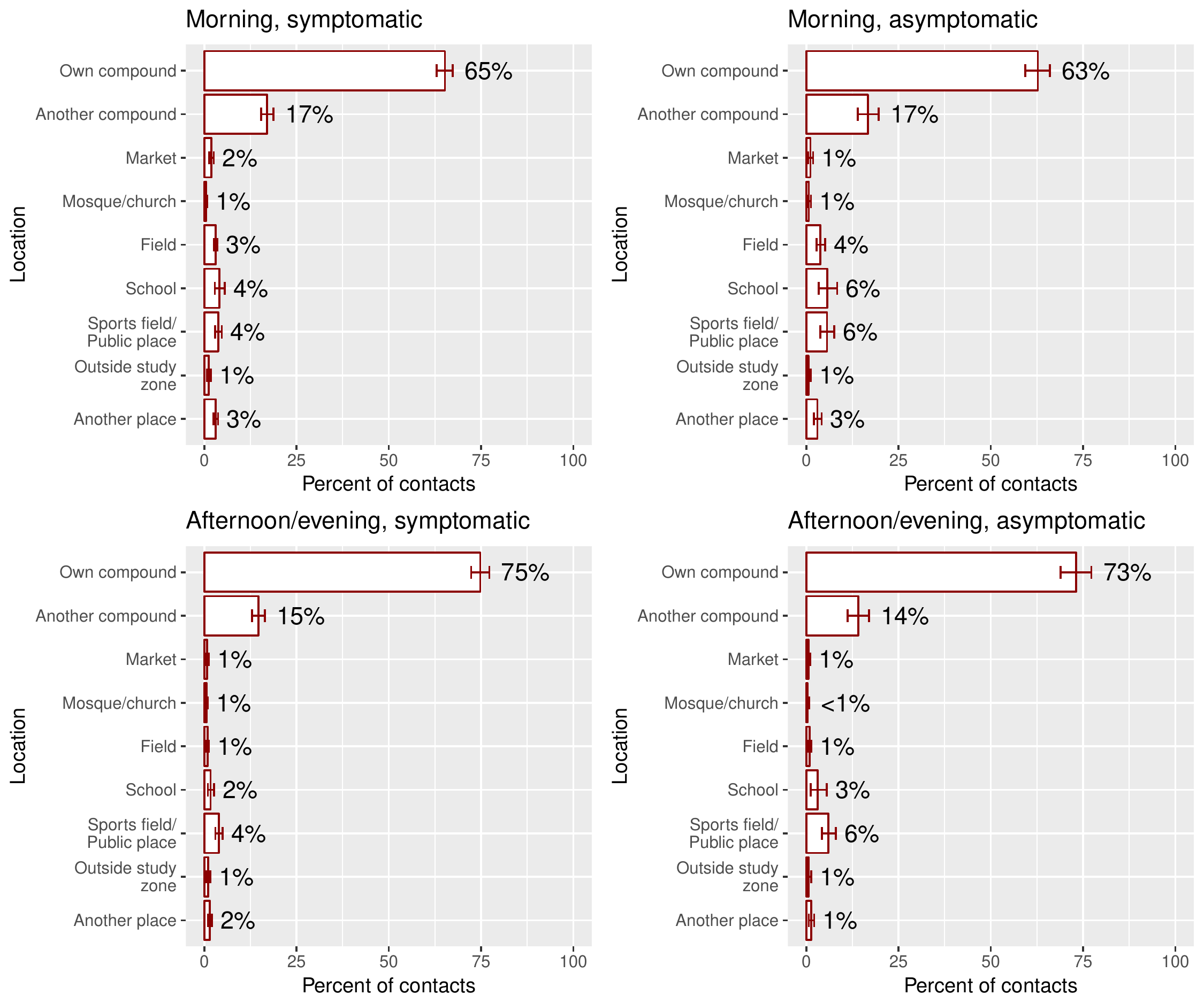} \end{figure}

We have demonstrated the reasoning for our update to the estimator assuming that all network members were surveyed.  When network members are randomly sampled, the rows of $S$ and columns of $V$ are sampled randomly, and the estimator is unbiased.  Our sampling process favored symptomatic people, who could be less likely to travel, but the data show no evidence for difference in travel patterns based on symptom status, as evidenced by Figure~\ref{fig:locsymp}.  This figure displays the location distribution of contacts by symptom status and time of day based on the multiply imputed data set.  Standard errors were calculated by generating 500 nonparametric bootstrap resamples of each imputed data set, pooling across the imputed data sets, and then calculating the 2.5\% and 97.5\% quantiles for each location proportion~\citep{schomaker2018bootstrap}.

\clearpage
\section{R code}
\begin{verbatim}
library(dplyr)
library(knitr)
library(xtable)
library(survival)
library(timereg)
  
# Function to compute change in cumulative hazard
# due to treatment over some number of years
cumulInc <- function(Y, inf.time, years, cens){
  Xt = solve(t(Y)%*%Y, t(Y))
  if(length(unique(time)) < length(time)) time[cens==0] <-  time[cens==0] + 
  runif(length(time[cens==0]),0,1)
  Xt = Xt[,inf.time<(years*365) & cens == 0]
  At = matrix(apply(Xt, 1, sum), nrow = 1)
  vcovmat = matrix(0, ncol = dim(Xt)[1], nrow = dim(Xt)[1])
  for(v in 1:dim(Xt)[2]){
    if(sum(is.na(Xt[,v]))==0)
      vcovmat = vcovmat + Xt[,v]%*%t(Xt[,v])
  }
  return(list(coefs = At, vcovmat = vcovmat))
}

#Function to update model formula with covariate terms
#including const() wrapper for time-invariance
update_form_const <- function(X, formula){
  if(is.null(dim(X))){
    return(update.formula(formula, .~. + const(X)))
  }else{
    if(is.null(names(X))){
      dimnames(X)[[2]] = paste0("X", 1:dim(X)[2])
      X = data.frame(X)
    }
    formTerms = terms(formula)
    modelTerms <- c(attr(formTerms, "term.labels"), paste0("const(", names(X), ")"))
    return(reformulate(modelTerms, response = attr(formTerms, "variables")[[2]]))
  }
}

## Function to fit additive hazards model
## returns results for randomized treatment effect
## and overall treatment effect.
fit_addHaz <- function(time, #time of event or censoring
                       cens, #indicator for censoring (1 = censored, 0 = event)
                       trt,  #randomized treatment assignment
                       X = NULL,   #matrix of additional covariates.  
                       # Assumed time-invariant
                       mix.pct, #percent of contacts to treated clusters
                       clust, #cluster membership
                       years, 
                       #time interval for difference in cumulative hazard
                       max.time=NULL, # end of follow-up for analysis
                       plot.trt = TRUE){  

  require(survival)
  require(timereg)
  

  # Resolve tied event times by adding draw from a uniform(0,1) distribution 

  set.seed(100)
  
  if(length(unique(time)) < length(time)) time[cens==0] <-  time[cens==0] + 
  runif(length(time[cens==0]),0,1)
  
  #set up survival data
	surv.data = Surv(time, 1-cens)
  
	#get randomized treatment effect estimate and SE
	if(is.null(X)){
	  form = surv.data ~ trt + cluster(clust)
	}else{
	  form = update_form_const(X, surv.data ~ trt + cluster(clust))
	}
	surv.fit = aalen(form, max.time=max.time)

	#get overall treatment effect estimate and SE
	Z = mix.pct
	if(is.null(X)){
	  form = surv.data ~ Z + cluster(clust)
	}else{
	  form = update_form_const(X, surv.data ~ Z + cluster(clust))
	}
	surv.fit.adj = aalen(form, max.time=max.time)

	if(plot.trt){
	  par(mfrow = c(1,2), mgp = c(2, 0.5, 0))
	  plot(surv.fit$cum[,1], surv.fit$cum[,3], type = "l", 
	  main = "No-contamination Estimator")
	  plot(surv.fit.adj$cum[,1], surv.fit.adj$cum[,3], type = "l", 
	  main = "Contamination-adjusted Estimator")
	} 
	
	#get change in cumulative incidence under randomized effect
  id = which(time == max(time[time <= years*365])) 
  
	id = which (surv.fit$cum[,1]==max(surv.fit$cum[,1]))[1]
  
  cumIncTrt = surv.fit$cum[id, 3]
  sd_CIT = surv.fit$robvar.cum[id, 3]  
  
  #get change in cumulative incidence under overall effect  	
  cumIncMix = surv.fit.adj$cum[id, 3]
  sd_CIM = surv.fit.adj$robvar.cum[id, 3]  
  
  return(list(rand_fit = surv.fit, 
         overall_fit = surv.fit.adj, 
         cumIncTrt = cumIncTrt, sd_CIT = sd_CIT,
         cumIncMix = cumIncMix, sd_CIM = sd_CIM))
}

# To get variance of incidence difference for covariates at values other than 0
# x1 (and x2 if wanting different values for treated (x1) and control (x2) for some reason)
# should be a vector of the same length as the covariates used in the model
incDiffX <- function(inc.res, x1, x2 = NULL){
  if(is.null(x2)) x2 <- x1
  
  sigma = inc.res$vcovmat
  coef1 = matrix(c(1,1,x1), nrow = 1)
  coef2 = matrix(c(1,0,x2), nrow = 1)
  
  sd_X <- sqrt(coef1%*%sigma%*%t(coef1)+coef2%*%sigma%*%t(coef2))
  return(sd_X)  
}





load('tte_dat')
load('intdat')
dat=left_join(tte_dat, intdat, by='village')
dat$censored = 1-dat$infected

mod=fit_addHaz(time=dat$tte, 
               cens=dat$censored, 
               trt=factor(dat$treatment),  
               X = NULL,   
               mix.pct=dat$pct, 
               clust=dat$village, 
               years=1,
               plot.trt = TRUE) 

load('tte_dat_SEASONAL')
dat=left_join(tte_dat, intdat, by='village')
dat$censored = 1-dat$infected

mod=fit_addHaz(time=dat$tte, 
               cens=dat$censored, 
               trt=factor(dat$treatment),  
               X = NULL,   
               mix.pct=dat$pct, 
               clust=dat$village, 
               years=1,
               plot.trt = TRUE) 

load('tte_dat_YEAR2')
dat=left_join(tte_dat, intdat, by='village')
dat$censored = 1-dat$infected

BEGIN = as.Date("7/15/2010", "%m/%d/%Y") # Beginning of follow-up for this year
strike.begin = as.Date("1/1/2011", "%m/%d/%Y")
max.time = strike.begin - BEGIN ## Censor analysis at beginning of strike 

mod=fit_addHaz(time=dat$tte, 
               cens=dat$censored, 
               trt=factor(dat$treatment),  
               X = NULL,   
               mix.pct=dat$pct, 
               clust=dat$village, 
               incYrs=1,
               max.time=max.time,
               plot.trt = TRUE) 
               
mod

# Exploratory analysis without censoring before strike
mod=fit_addHaz(time=dat$tte, 
               cens=dat$censored, 
               trt=factor(dat$treatment),  
               X = NULL,   
               mix.pct=dat$pct, 
               clust=dat$village, 
               years=1,
               max.time=
               plot.trt = TRUE) 
\end{verbatim}

\section{Data Cleaning}

Village of residence was recorded during  quarterly censuses of the Niakhar population by the Niahkar Demographic Surveillance System.~\cite{delaunay2002niakhar}  If participants moved during the trial, their departure date, arrival date, and village of their new residence were recorded.  Those who moved a second time had their departure date (but not residence after second move) recorded as well.  The cleaning process for inconsistencies in the recorded movement data is described below:

\begin{enumerate}
\item	In 8 cases, the departure and arrival dates of the second move were earlier than those of the first move.  For these cases, the information for second and first moves was swapped.
\item	In 46 cases, the arrival date of the second move was earlier than the arrival and departure dates of the first move, and the departure date of the second move was missing.  For these cases, the information for second and first moves was swapped.  After the swap, the (missing) departure date for the first move was imputed to be the arrival date of the second.
\item	In 13 cases where the departure date of the first move was missing, it was imputed to be the arrival date of the second move.
\item	In 83 cases where the arrival date of the second move was earlier  than the departure date of the first, the departure date of the first was recoded to equal the arrival date of the second.
\item	In 13 cases where the departure date of the first move was earlier than the arrival date of the first move, and the arrival date of the second move was non-missing, the departure date of the first move was recorded to be the arrival date of the second.
\item	After these updates were made, there were 13 cases that did not have arrival and departure dates in sequential order (i.e., arrival 1 $\le$ departure 1 $\le$ arrival 2 $\le$ departure 2); these were excluded from analysis.
\item The movement data was recorded by storing the village, arrival date, and departure date, of the first ``stay'' and the second ``stay'', as well as an overall ``village'' variable.  In over 99\% of cases, the village variable matched that of the first stay.  However, there were 168 participants for whom the overall village variable differed from that of the first stay.  This is because movement data were recorded differently for this small subset of the data.  For them ``village'' indicated the village of residence prior to the first stay rather than the village of the first stay.  As such, these cases had up to three distinct residence stays recorded, which differs from the rest of the data which only had up to two distinct stays recorded.  These cases were re-coded to be consistent with the rest of the data by transferring the village information (which actually describes the first distinct stay) to the variables for the first stay (so that village and village.stay.1 are consistent), transferring the information recorded for the first stay (actually the second stay) to the variables for the second, and removing information for the second (actually third) stay, as follows:

\begin{enumerate}
\item  village.stay.1 was  re-coded to village 
\item  arrival.date.stay.1 was  re-coded to birth.date 
\item   departure.date.stay.1 was re-coded to arrival.date.stay.1 
\item  village.stay.2 was re-coded to village.stay.1 
\item  arrival.date.stay.2 was re-coded to the original  arrival.date.stay.1 (the updated departure.date.stay.1)
\item  departure.date.stay.2 was re-coded to departure.date.stay.1 
\end{enumerate}

\end{enumerate}

\end{document}